\documentstyle[aps,epsf,rotate,multicol]{revtex}

\begin{document}

\draft

\title{Universal features in the growth dynamics of complex
organizations}


\author{Youngki Lee$^1$, Lu\'{\i}s~A.~Nunes Amaral$^{1,2}$, David 
Canning$^3$, Martin Meyer$^1$, and H.~Eugene Stanley$^1$}

\address{
$^1$Center for Polymer Studies and Department of Physics,
  	Boston University, Boston, MA 02215 \\
$^2$Department of Physics, Massachusetts Institute of Technology,
	Cambridge, MA 02139 \\
$^3$Harvard Institute for International Development, Harvard
	University, Cambridge, MA 02136 \\ 
}


\maketitle

\begin{abstract}

We analyze the fluctuations in the gross domestic product (GDP) of 152
countries for the period 1950--1992.  We find that (i) the
distribution of annual growth rates for countries of a given GDP
decays with ``fatter'' tails than for a Gaussian, and (ii) the width
of the distribution scales as a power law of GDP with a scaling
exponent $\beta \approx 0.15$.  Both findings are in surprising
agreement with results on firm growth.  These results are consistent
with the hypothesis that the evolution of organizations with complex
structure is governed by similar growth mechanisms.

\end{abstract}

\begin{multicols}{2}

  In the study of physical systems, the analysis of the scaling
properties of the fluctuations has been shown to give important
information regarding the underlying processes responsible for the
observed macroscopic behavior.  In contrast, most studies on the time
evolution of economic time series have concentrated on average growth
rates
\cite{Dowrick89,Barro91,Mankiw92,Head95,Durlauf95,Durlauf96,Sala96,
Quah96,Bernard96,Galor96,Gibrat33,Hart,Simon,Evans,Hall,Davis,Radner,
Sutton97}.  Here, we investigate the possibility that the study of
fluctuations in economics may also lead to a better understanding of
the mechanisms responsible for the observed dynamics
\cite{Mandelbrot,Bak,Bouchaud,Levy,Durlauf9x}.

  We therefore analyze the fluctuations in the growth rate of the
gross domestic product (GDP) of 152 countries during the period
1950--1992 \cite{Summers91}.  We will show that (i) the distribution
of annual growth rates for countries of a given GDP is consistent for
a certain range with an exponential decay, and (ii) the width of the
distribution scales as a power law of GDP with a scaling exponent
$\beta \approx 0.15$.  Both findings are in surprising agreement with
results reported on the growth of firms
\cite{Stanley96,Amaral97a,Buldyrev97}.

  It is not obvious that firms and countries show similarities other
than that they are complex systems made up of interacting
individuals. Hence, our findings raise the intriguing possibility that
similar mechanisms are responsible for the observed growth dynamics
of, at least, two complex organizations: firms and countries.

\begin{figure}
\narrowtext
\centerline{
\epsfysize=0.9\columnwidth{\rotate[r]{\epsfbox{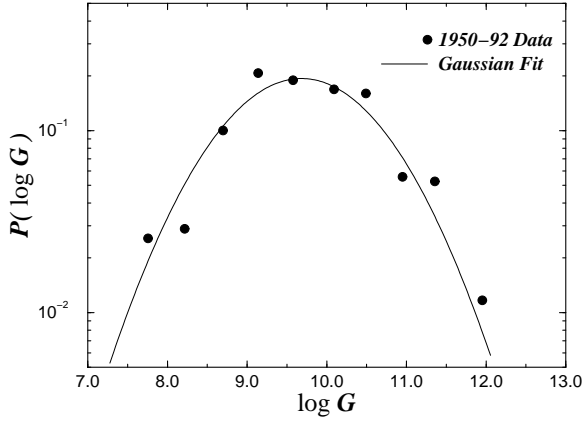}}}}
\vspace*{0.5cm}
\caption{ Probability distribution of the logarithm of GDP. The data
have been detrended by the average growth rate, so values for
different years are comparable. The data points are the average over
the entire period, '50-'92, and the continuous line is a Gaussian fit
to the data. We also confirmed that the distribution is stationary ---
i.e., remains the same for different time intervals. }
\label{f-gdp}
\end{figure}

  We first study the distribution $p(\log G)$, where G is the value of
the GDP detrended by the global average growth rate, for all the
countries and years in our database.  As shown in Fig.~\ref{f-gdp},
$p(\log G)$ is consistent with a Gaussian distribution, implying that
$P(G)$ is log-normal.  We also find that the distribution $P(G)$ does
not depend on the time period studied.

\begin{figure}
\narrowtext
\centerline{
\epsfysize=0.9\columnwidth{\rotate[r]{\epsfbox{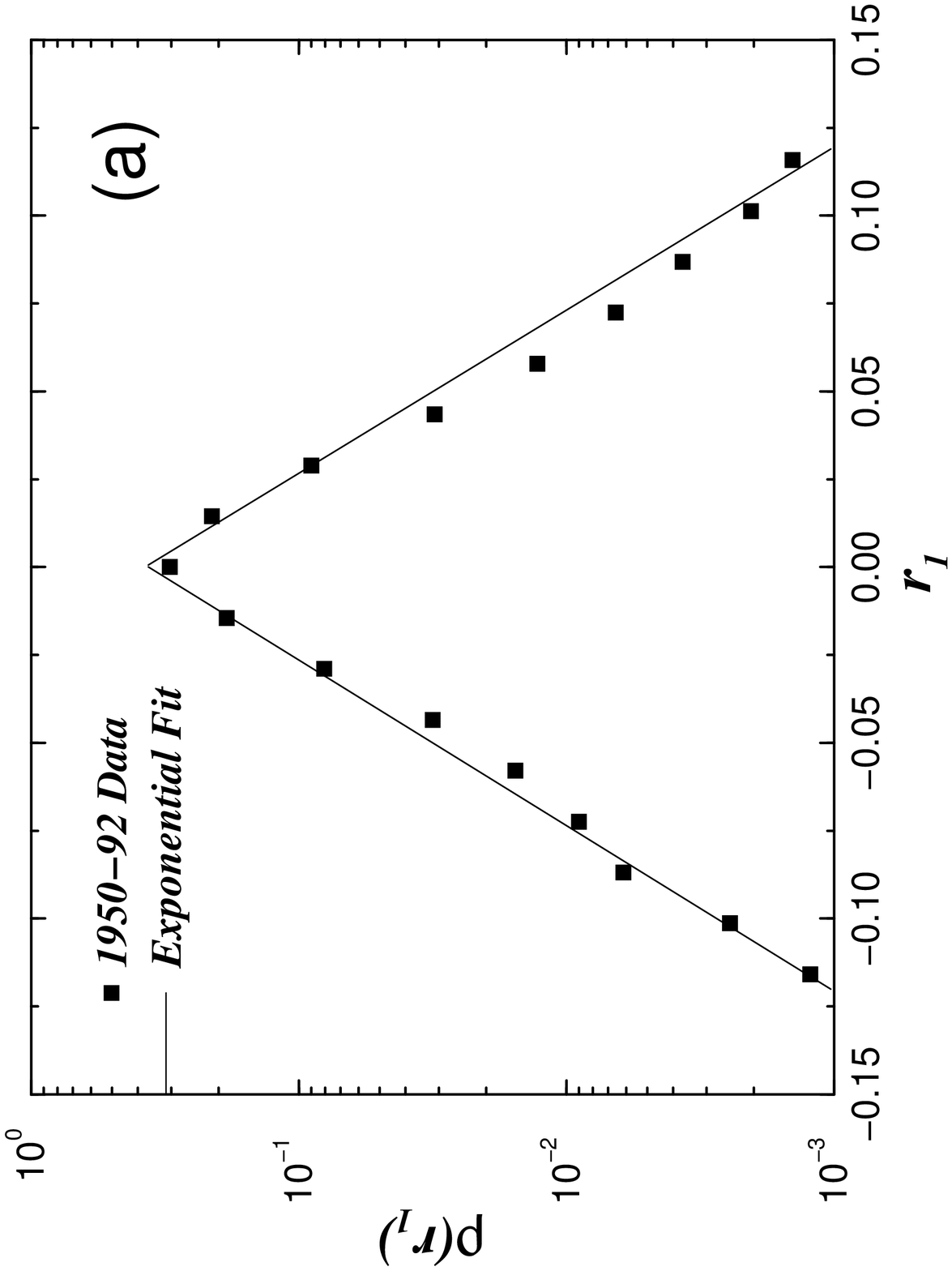}}}}
\vspace*{0.5cm}
\centerline{
\epsfysize=0.9\columnwidth{\rotate[r]{\epsfbox{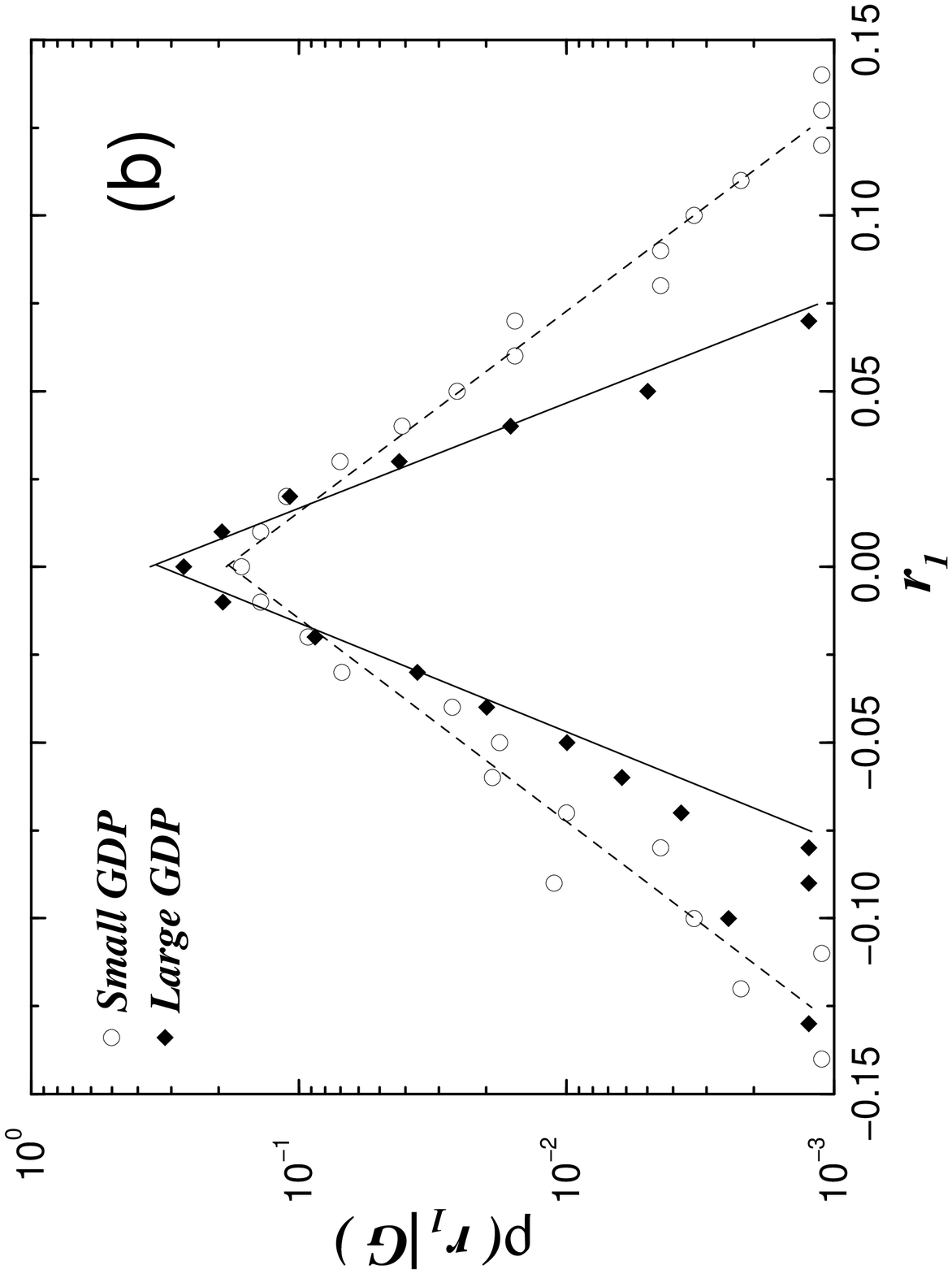}}}}
\vspace*{0.5cm}
\caption{(a) Probability density function of annual growth rate $r_1$.
Shown are the average annual growth rates for the entire period
1950--1992 together with an exponential fit, as indicated in
Eq.~(1). (b) Probability density function of annual growth rate for
two subgroups with different ranges of $G$, where $G$ denotes the GDP
detrended by the average yearly growth rate.  The entire database was
divided into three groups: $6.9\times10^7\le G < 2.4\times10^9$,
$2.4\times10^9\le G < 2.2\times10^{10}$, and $2.2\times10^{10} \le G <
7.6\times10^{11}$, and the figure shows the distributions for the
smallest and largest groups.  }
\label{f-rho}
\end{figure}

  Next, we calculate the distribution of annual growth rate $r_1
\equiv \log( G(t+1) / G(t))$, where $G(t)$ and $G(t+1)$ are the GDP of
a country in the years $t$ and $t+1$. In the limit of small annual
changes in $G$, $r_1(t)$ is the {\it relative\/} change in $G$.  For
all countries and all years, we find that the probability density of
$r_1$ is consistent, for a certain range of $|r_1|$, with an
exponential decay (see Fig.~\ref{f-rho}a)
\begin{equation}
\rho(r_1) = \frac{1}{\sqrt{2}\sigma_o} \exp\left(-{\sqrt{2}\,|r_1 - \bar
r_1|\over\sigma_o}\right)\,, \label{e-dist0} 
\end{equation}
where $\sigma_o$ is the standard deviation.  We find that the
functional form of the distribution is stable over the entire period
considered, i.e. we find the same distribution for all time intervals.

  We then investigate how the growth rate distribution depends on the
initial value of the GDP.  Therefore, we divide the countries into
groups according to their GDP.  We find that the empirical {\it
conditional\/} probability density of $r_1$ for countries with
approximately the same GDP is also consistent in a given range with
the exponential form (see Fig.~\ref{f-rho}b)
\begin{equation}
\rho(r_1|G) = \frac{1}{\sqrt{2}\sigma(G)}
\exp\left(-{\sqrt{2}\,|r_1 - \bar r_1|\over\sigma(G)}\right)\,,
\label{e-distribution}
\end{equation}
where $\sigma(G)$ is the standard deviation for countries with GDP
equal to $G$.  Using a saddle point approximation, we may integrate
the distribution (\ref{e-distribution}) over $P(G)$ using a log-normal
distribution and recover (\ref{e-dist0}).

\begin{figure}
\centerline{
\epsfysize=0.9\columnwidth{\rotate[r]{\epsfbox{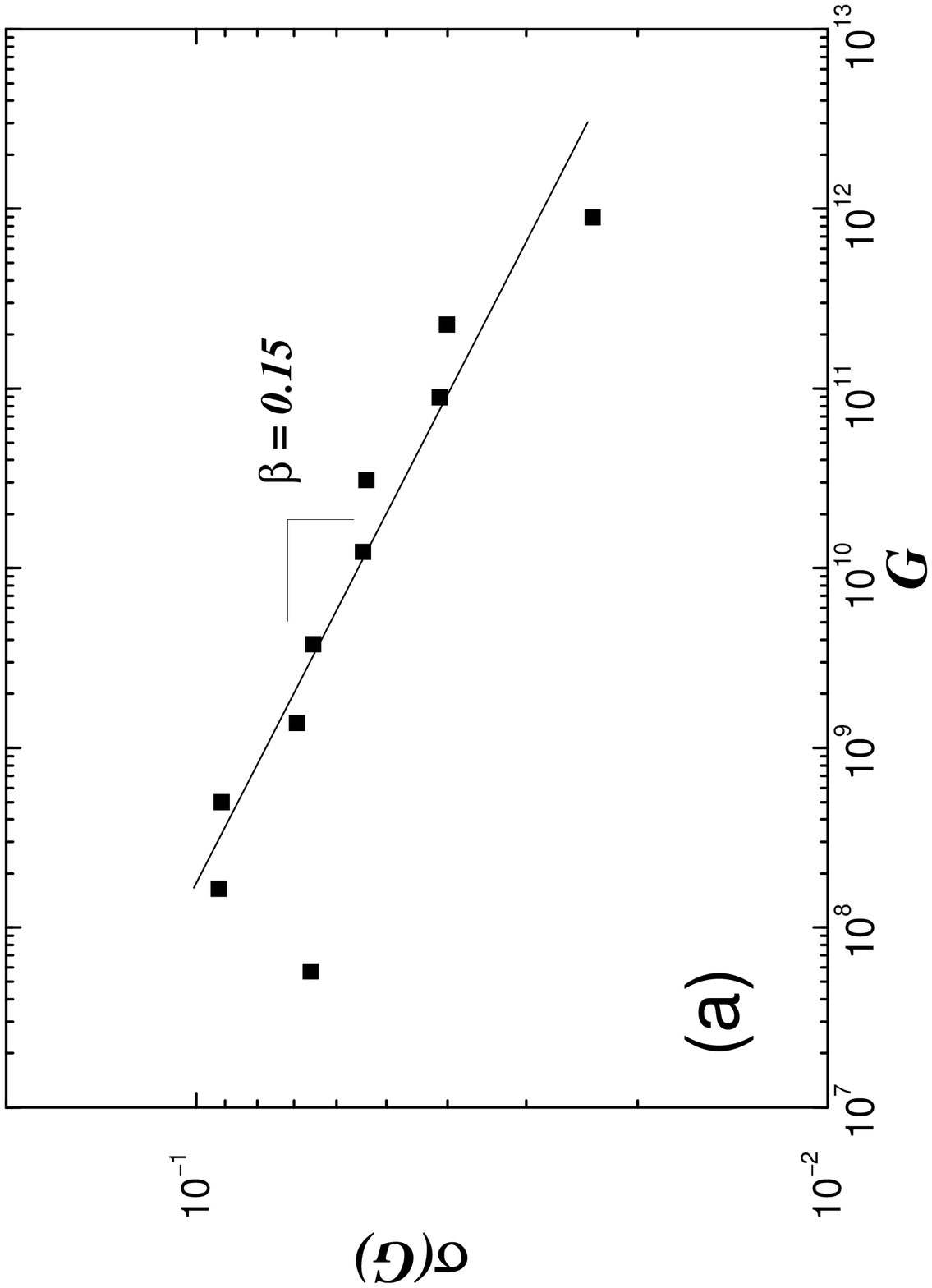}}}}
\vspace*{0.5cm}
\centerline{
\epsfysize=0.9\columnwidth{\rotate[r]{\epsfbox{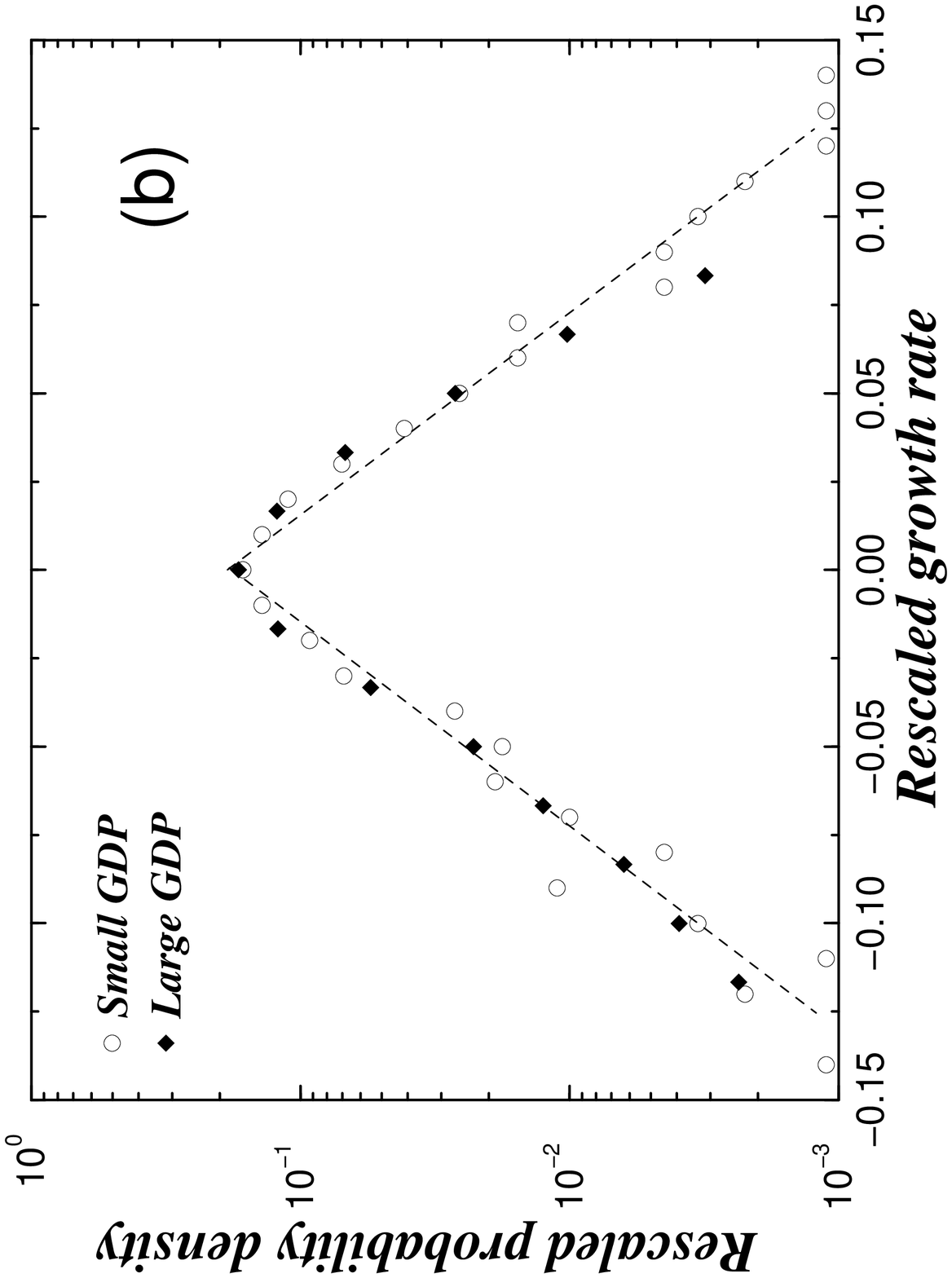}}}}
\vspace*{0.5cm}
\caption{(a) Plot of the standard deviation $\sigma(G)$ of the
distribution of annual growth rates as a function of $G$, together
with a power law fit (obtained by a least square linear fit to the
logarithm of $\sigma$ vs the logarithm of $G$).  The slope of the line
gives the exponent $\beta$, with $\beta=0.15$. (b) Rescaled
probability density function, $\sigma(G)\rho(r_1|G)$, of the rescaled
annual growth rate, $r_1 / \sigma(G)$. Note that all data collapse
onto a single curve. }
\label{f-sigma}
\end{figure}

  Figure~\ref{f-sigma}a shows that $\sigma(G)$ scales as a power law
\begin{equation}
\sigma(G) \sim G^{-\beta},
\label{e-sigma}
\end{equation}
with $\beta \approx 0.15$.  We confirm our results by a
maximum-likelihood analysis \cite{NumRep}. In particular, we find that
the log-likelihood of $\rho(r_1|G)$ being described by an exponential
distribution --- as opposed to a Gaussian distribution --- is of the
order of $e^{600}$ to $1$.

\begin{figure}
\narrowtext
\centerline{
\epsfysize=0.9\columnwidth{\rotate[r]{\epsfbox{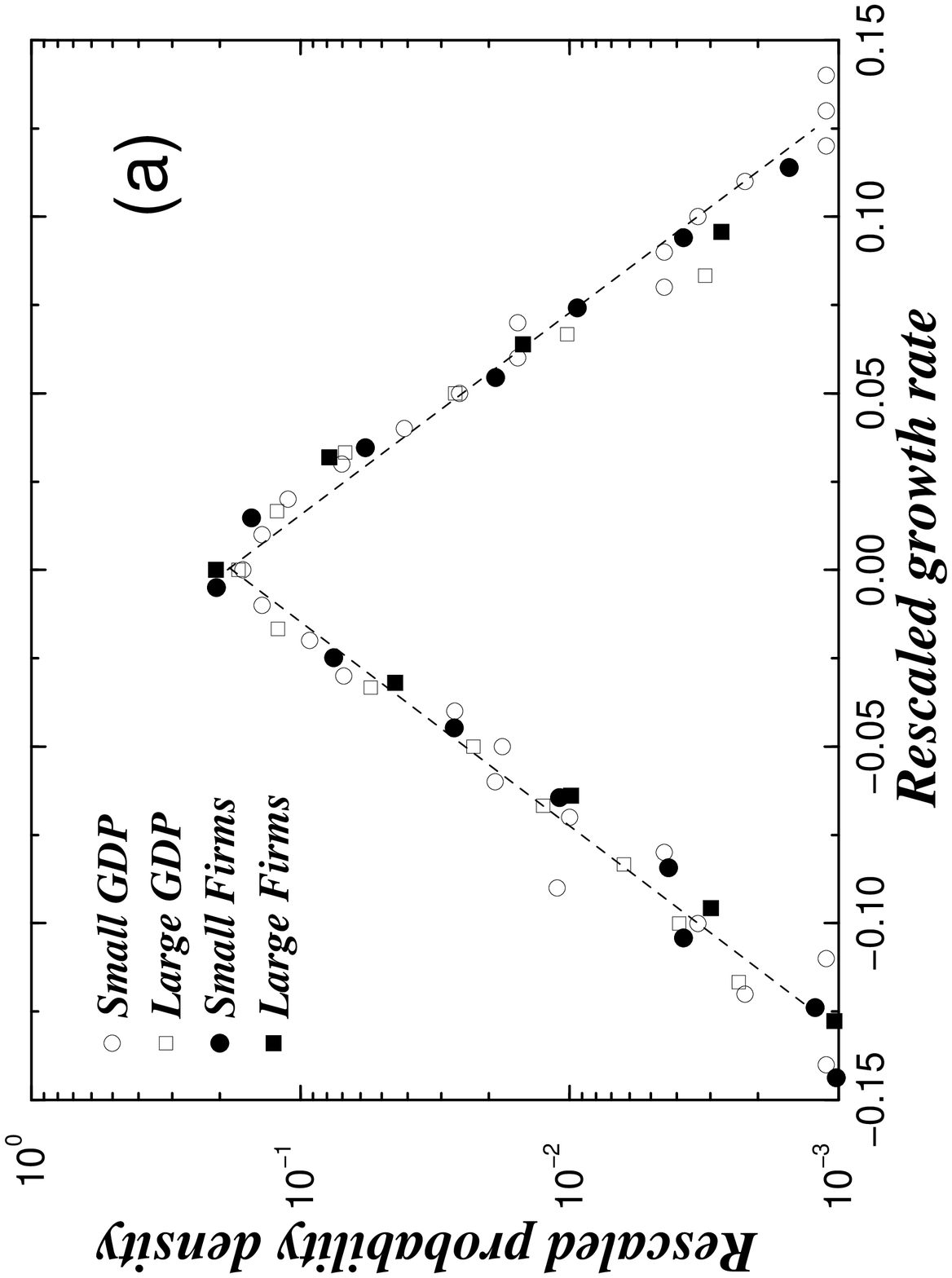}}}}
\vspace*{1.5cm}
\centerline{
\epsfysize=0.9\columnwidth{\rotate[r]{\epsfbox{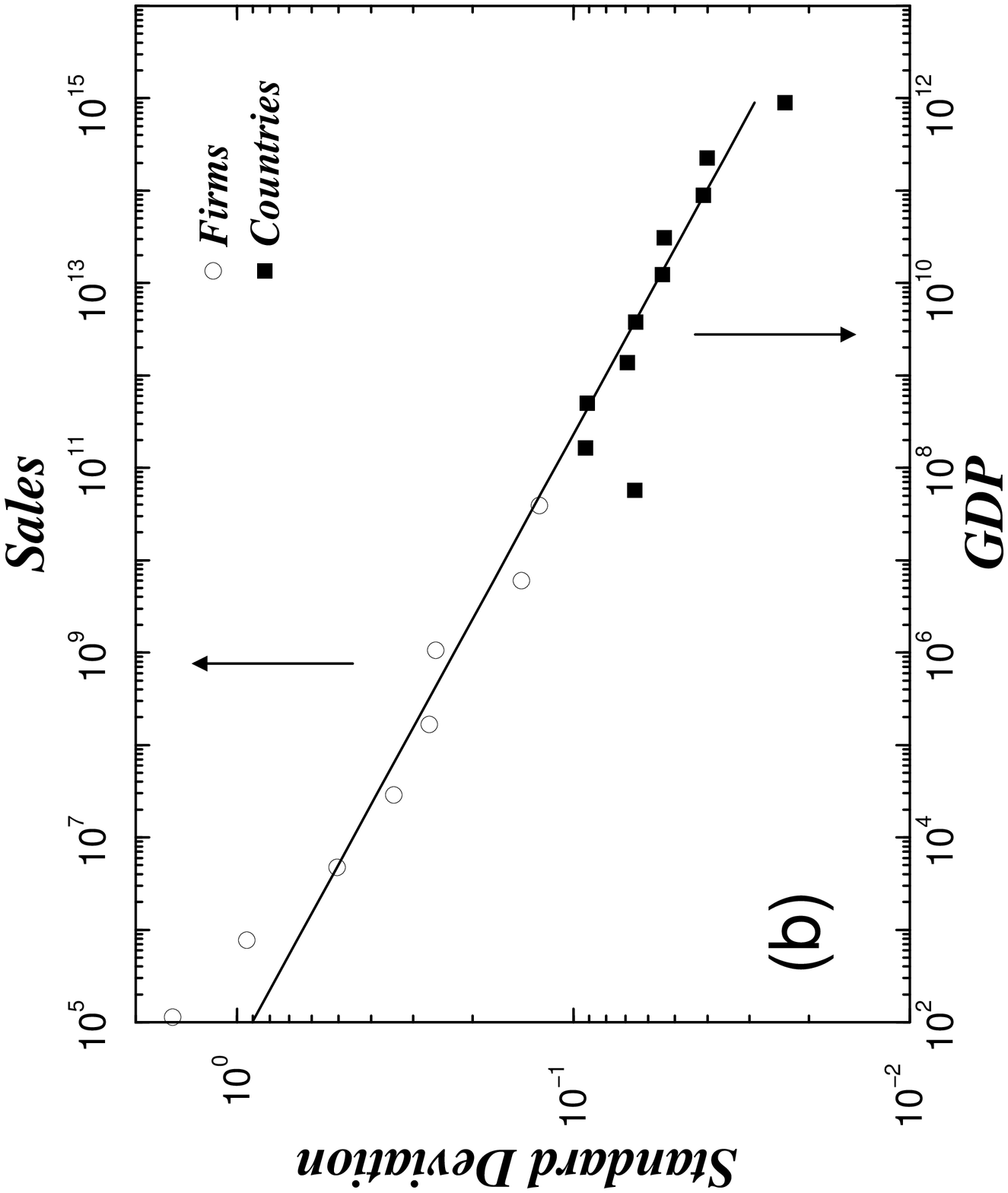}}}}
\vspace*{0.5cm}
\caption{ Test of the similarity of the results for the growth of
countries and firms. (a) Conditional probability density of annual
growth rates for countries and firms.  We rescale the distributions as
in Fig.~3b.  All data collapse onto a single curve showing that indeed
the distributions have the same functional form. (b) Standard
deviation of the distribution of annual growth rates. Note that
$\sigma$ decays with size with the {\it same\/} exponent for both
countries and firms. The firm data were taken from the Compustat
database for publicly-traded manufacturing firms from 1974-1993 (see
\protect\cite{Amaral97a} for details).}
\label{f-univ}
\end{figure}

  The results of Figs.\ref{f-gdp}-\ref{f-sigma} are in {\it
quantitative\/} agreement with findings for the growth of firms
\cite{Stanley96,Amaral97a,Buldyrev97}.  Figure~\ref{f-univ}a shows
that the same functional form describes the probability distribution
of annual growth rates for both the GDP of countries and the sales of
firms. Moreover, as shown in Fig.~\ref{f-univ}b, the width of the
distribution of annual growth rates also decays with size with the
same exponent for firms and countries.

  The fact that the {\it same\/} empirical laws hold for the growth
dynamics of both countries and firms suggests that a {\it common\/}
mechanism applies to both processes. To explore this possibility, we
consider two limiting models.  

	$(i)$ Assume that an economic organization, such as a country
or a firm, is made up of many units, which are of identical size and
grow independently of one another. Then, the growth fluctuations as a
function of size decay as a power law with an exponent
$\beta=0.5$. This result is due to the fact that the number of units
forming a given organization is proportional to its size, and because
the variance of the sum of $n$ independent quantities grows like
$\sqrt{n}$ \cite{Amaral97a}. 

	$(ii)$ Assume that there are very strong correlations between
the units, which is the opposite limiting case.  Then, it follows that
the growth dynamics are indistinguishable from the dynamics of
structureless organizations.  As a result, we obtain an exponent
$\beta = 0$, i.e., there is no size dependence of $\sigma$.

  The fact that the exponent $\beta$ for the empirical data is in
between the two limiting cases shows that the models $(i)$ and $(ii)$
are both based on false assumptions.
Our results are consistent with a recently proposed model
\cite{Amaral97b} for the growth of organizations.  The dynamics of the
model give rise to subunits whose characteristic size increases with
the size of the organization leading to an exponent $\beta$ smaller
than $1/2$.


  Our empirical results suggest an important consequence for economic
growth: Although large economies tend to diversify into a wider range
of economic activities leading to smaller relative fluctuations, the
degree of diversification observed is much smaller than what would be
expected if diversification would increase linearly with the size of
the economy --- which would correspond to $\beta = 0.5$. This effect
is quantitatively the same for firms and countries, which raises the
intriguing possibility that a {\it common\/} mechanism might
characterize the growth dynamics of economic organizations with
complex internal structure. The existence of ``universal'' mechanisms,
which can give rise to general laws that are independent of the
particular details of the system, could provide a firmer grounding for
the application of physics methods to questions in economics
\cite{Mandelbrot,Bak,Bouchaud,Levy,Durlauf9x}.

  We thank E.~Alexander, S.~Alexander, S.V. Buldyrev, Y.~Liu,
P.~Gopikrishnan, V.~Horv\'ath, C.-K. Peng, M.A. Salinger, and
especially S.~Havlin and J.D.~Sachs for helpful discussions.  Y.L.\
thanks BP and the Korea Research Foundation for support, L.A.N.A.\
thanks the JNICT, and M.M.\ thanks the DFG.  The CPS is supported by
grants from NSF and NIH.

\end{multicols}

\end{document}